\documentclass[prl,twocolumn,showpacs,preprintnumbers,amsmath,amssymb]{revtex4}

\newcommand{\beq}{\begin{equation}}

\newcommand{\eeq}{\end{equation}}
\newcommand{\myref}[1]{~{(\ref{#1})}}
\newcommand{\mycite}[1]{~{\cite{#1}}}

\newcommand{\tr}{\mathrm{tr}}

\def \be  {\begin{equation}}
\def \ee  {\end{equation}}
\def \ba  {\begin{eqnarray}}
\def \ea  {\end{eqnarray}}

\usepackage{graphicx}
\usepackage{dcolumn}
\usepackage{bm}

\begin{document}

\preprint{ITEP-TH-51/08}

\title{Nonlocal Gluon Condensate  from the Dyson--Schwinger Equations}

\author{A.~V.~Zayakin$^{\flat\ddagger}$}
 \email{Andrey.Zayakin@physik.lmu.de}
\author{J.~Rafelski$^{\flat\sharp}$} \email{rafelski@physics.arizona.edu}

\affiliation{$^{\flat}$
Department f\"ur Physik der Ludwig-Maximillians-Universit\"at M\"unchen und\\
 Maier-Leibniz-Laboratory,Am Coulombwall 1, 85748 Garching, Germany
}
\vspace{2mm}
\affiliation
{$^{\ddagger}$
ITEP, B.Cheremushkinskaya, 25, 117218, Moscow, Russia
}
\vspace{2mm}
\affiliation{$^\sharp$ Department of Physics, University of Arizona, Tucson, Arizona, 85721 USA}

\date{\today}

\begin{abstract}
We establish a novel method of obtaining the gauge-invariant bilocal condensate of gluon fields from the solutions of the Dyson--Schwinger equations (DSE) which works in the infrared and ultraviolet limits.
We present explicit results for  $SU(3)$ case in four dimensions, and compare to other related analytical and lattice results.
\end{abstract}
\maketitle
{\bf Motivation:}
The non-perturbative QCD vacuum is known for the presence of gluon (and quark) condensates\mycite{Shifman:1978bx}. These objects are in general studied in a local approximation. A bilocal object can be introduced by using operator product expansion (OPE)\mycite{Mikhailov:1992ug}. However, OPE method can work only at a short scale.  The infrared~(IR) behaviour of gluon condensate is of importance because it acts as a potential in which quarks fluctuate\mycite{Lavelle:1988eg}. Therefore, to understand, e.g. how QCD vacuum properties are modified in the infrared by external electromagnetic fields\mycite{Rafelski:1998tc}, one must first understand the dynamics of the gauge field condensate and than quark condensate in a scheme operational in the infrared.

We obtain here   the gauge-invariant nonlocal condensate in terms of perturbative and non-perturbative propagators. Then we solve numerically the Dyson--Schwinger equations (DSE) for $SU(3)$ gauge theory at approximately two loops, following the methods developed by Bloch\mycite{Bloch:2003yu} and other authors to obtain the non-perturbative propagators. We introduce perturbative propagators in a way which is consistent with the non-perturbative ones, and extract the gauge field condensate. Finally we analyze the condensate dependence on distance, and compare with prior results obtained by analytical and lattice methods.

{\bf Form factors and nonlocal condensate:}
We start with the $SU(N_c=3)$ gauge field action
\beq
S=-\frac{1}{4}\frac{\tr}{N_c} \int d^4 x G_{\mu\nu} G^{\mu\nu},
\eeq
where $G_{\mu\nu}=G_{\mu\nu}^aT^a$, $G_{\mu\nu}^a=\partial_\mu A_\nu^a-\partial_\nu A_\mu^a +gf^{abc}A_\mu^b A_\nu^c$, $\tr T^a T^b=N_c \delta^{ab}$.
The gluon propagator $D_{F\mu\nu}^{ab}$   is considered in Landau gauge
$D_{F\mu\nu}^{ab}(x)=-i\langle T\left(A_\mu^a(x) A_\nu^b(0)\right)\rangle$
with a form factor $F(p^2)$ 
\beq\label{DForm}
D_{F\mu\nu}^{ab}(p) =
\delta^{ab}\left(g_{\mu\nu}-\frac{p_\mu p_\nu}{p^2}\right)\frac{F(p^2)}{p^2+i\epsilon},
\eeq
and the ghost propagator is defined as $D^{Gab}(x)=-i\langle T\left(c^a(x) c^b(0)\right)\rangle$,
with the ghost form factor $G(p^2)$,
\beq
D^{Gab}(p)=\frac{\delta^{ab}}{p^2+i\epsilon}G(p^2).
\eeq

Following Ref.\mycite{Dosch:1988ha}, the bilocal condensate is defined as the vacuum expectation value (VEV) of the following gauge-invariant operator
\beq\label{bilocal}\displaystyle
\mathcal{F}(y-z)=\langle :\frac{\alpha}{\pi}\frac{\tr}{N_c} \left[G_{\mu\nu}(y) W(y,z) G^{\mu\nu} (z) W(z,y) \right]:\rangle,
\eeq
where Wilson--Polyakov lines are $W(y,z)=\mathcal{P}\mathrm{exp}\left(ig\int^z_y A_\mu dx^\mu\right)$, $\mathcal{P}$ being the path-ordering symbol. Note that the normal ordering in Eq.\myref{bilocal} is defined with respect to the perturbative (PT) vacuum, whereas the operators are averaged over a non-perturbative (NP) vacuum.

We are going to work at below  the two-loop limit, thus Eq.\myref{bilocal} is expanded perturbatively up to $\alpha^2$-order:
\beq
\mathcal{F}=\mathcal{F}^{(0)}+\mathcal{F}^{(1)}_{1}+
\mathcal{F}^{(1)}_{2}+\mathcal{F}^{(1)}_{3}
\eeq
indices $^{(0)},^{(1)}$ denoting $\alpha$ order.
The two-gluon piece is
\beq\label{twogluon}
\mathcal{F}^{(0)}=\langle:\frac{\alpha}{\pi} \frac{\tr}{N_c} \left[ G^{\mathrm{Lin}\mu\nu}(x) G^{\mathrm{Lin}}_{\mu\nu}(0)\right]:\rangle,
\eeq
where we have split the gauge covariant quantity $G_{\mu\nu}$ into a linear (Abelian) and a bilinear  non-gauge covariant objects $G^{\mathrm{Lin}}_{\mu\nu}=G_{\mu\nu}+i g[A_\mu,A_\nu].$
Three-gluon term is absent, and a four-gluon term without Wilson lines is
\beq
\mathcal{F}^{(1)}_{1}=\langle :
\frac{\alpha}{\pi}\frac{\tr}{N_c} \left\{g[A_\mu(x),A_\nu(x)] g[A_\mu(0),A_\nu(0)] \right\}:
\rangle.
\eeq
Wilson lines do not non-trivially contribute to the above $\mathcal{F}^{(0)}$ and $\mathcal{F}^{(1)}_{1}$, but contribute to  $\mathcal{F}^{(1)}_{2}$ and
$\mathcal{F}^{(1)}_{3}$:
\beq\displaystyle
\begin{array}{rcl}\displaystyle
\mathcal{F}^{(1)}_{2}(x)\!&\!\displaystyle=\!&\!
\displaystyle \langle :\frac{\alpha}{\pi}
\frac{\tr}{N_c}  \left[ \int\!\!\!\!\int g^2 F_{\mu\nu}(x) A_\beta\left(u(\tau)\right)
A_\gamma\left(v(\sigma)\right)\times\right.\\ \displaystyle
&\times&\displaystyle F^{\mu\nu}(0) \dot{u}^\beta \dot{v}^\gamma  \theta(\tau-\sigma)d\tau d\sigma \bigg]:
\rangle,\\ \displaystyle
\mathcal{F}^{(1)}_{3}\!&\!\displaystyle=\!&\!
\displaystyle\langle :\frac{\alpha}{\pi}
\frac{\tr}{N_c} \left[ \int\!\!\!\!\int g^2 F_{\mu\nu}(x) A_\beta\left(u(\tau)\right) \times\right.\\ &\times&\displaystyle F^{\mu\nu}(0) A_\gamma\left(v(\sigma)\right)\dot{v}^\gamma \dot{u}^\beta d\tau d\sigma \bigg]:
\rangle
\end{array}
\eeq
We use here straight lines $v^\gamma(\sigma)=\sigma x^\gamma,u^\beta(\tau)=\tau x^\beta,$
running over $\tau\in (0,1),\sigma\in (0,1)$.

To express $\mathcal{F}^{(0)},\mathcal{F}^{(1)}_{1,2,3}$ in terms of propagators, we need to handle the PT-normal products in the NP-vacuum. Consider  an  operator $\mathcal{O}(x) $. A $T$-product of operators can be evaluated in both perturbative and non-perturbative vacuum, with $G^{PT,NP}_\mathcal{O}$ defined as follows:
\beq
\begin{array}{rcl}
T\left(\mathcal{O}(x) \mathcal{O}(y)\right)&=&:\mathcal{O}(x) \mathcal{O}(y):^{PT}+iG^{PT}_\mathcal{O}(x,y)= \\[.2cm]
&=&
:\mathcal{O}(x) \mathcal{O}(y):^{NP}+iG^{NP}_\mathcal{O}(x,y),
\end{array}
\eeq
indices $NP$ and $PT$ over normal ordering symbols and Green's functions meaning the definitions of the latter in the nonperturbative or perturbative vacuum correspondingly. Taking a VEV over the nonperturbative vacuum, and subtracting the $NP$ definition of $T\left[\mathcal{O}(x) \mathcal{O}(y)\right]$ from the $PT$ one, one sees that
\beq
\langle :\mathcal{O}(x) \mathcal{O}(y):^{PT}\rangle_{NP}=
i(G^{NP}_\mathcal{O}(x,y)-G^{PT}_\mathcal{O}(x,y)).
\eeq
Thus emerges the object of interest
\beq
\Delta_F(p^2) \equiv \alpha(p^2)\left(F^{NP}(p^2)-F^{PT}(p^2)\right).
\eeq

In our case the non-perturbative form factor $F^{NP}(p^2)=F(p^2)$, Eq.\myref{DForm}, is what we obtain from solving the DSE. The corresponding perturbative form factor $F^{PT}(p^2)$ is discussed below. For the sake of generality we present, using Eq.\myref{twogluon}, the components of condensates in an arbitrary dimension $d$:
\beq
\mathcal{F}^{(0)}=2(d-1)d_c \int \frac{d^d p}{(2\pi)^d} \Delta_F(p^2) e^{ipx},
\eeq
in agreement with Ref.\mycite{Namyslowski:1991wu}, where a local limit of this equation is given, here and below $d_c=N_c^2-1$. For the subleading terms we obtain:
\beq
\mathcal{F}^{(1)}_1=4 d_c N_c\left[(d^2-2d-1) I_1^2-I_{2\mu\nu} I_2^{\mu\nu}\right],
\eeq
where the propagator moments are defined as
\beq
\begin{array}{rcl}
I_1(x)&=&\int \frac{d^d p}{(2\pi)^d}\frac{\Delta_F(p^2)}{p^2}e^{ipx} \\[.2cm]
I_{2\mu\nu}(x)&=&\int \frac{d^d p}{(2\pi)^d}\frac{\Delta_F(p^2)}{p^2} \frac{ p_\mu p_\nu}{p^2}e^{ipx}.
\end{array}
\eeq
For the terms containing Wilson lines, one has
\beq
\mathcal{F}^{(1)}_2+\mathcal{F}^{(1)}_3=4 d_c N_c W_1 W_2,
\eeq
the moments $W_{1},W_{2}$ being
\begin{eqnarray}
W_{1}\!&=&\int \frac{d^d p}{(2\pi)^4}\Delta_F(p^2)e^{ipx},\\[.2cm] \notag
W_{2}\!&=& \!12(1-d)\!\int\! \!\frac{d^d p}{(2\pi)^d}\frac{\Delta_F(p^2)}{p^2}
\!\left(x^2\!-\frac{(px)^2}{p^2}\right)\!\frac{\sin^2 \left(\frac{px}{2}\right)}{(px)^2}.
\end{eqnarray}
To be able to complete the evaluation of the nonlocal condensate we should supply the non-perturbative and perturbative form factors into these equations.

{\bf Dyson--Schwinger equations:}
To obtain the needed non-perturbative input form factor $F^{NP}(p^2)$  one option is to solve DSE. We work in pure gluodynamics, following  closely Bloch's method\mycite{Bloch:2003yu} of IR analysis and his truncation labeled (A.1), Table 3 ibid., thus we do not give here the full details. The procedure of writing down, truncating, IR-analyzing and solving DSE is   developed and described in detail in many works, see for example\mycite{Hauck:1996sm,Fischer:2003rp, Fischer:2008ic,Fischer:2008yv,
Alkofer:2008bs}. The DSE equations in a covariant gauge are organized in the following way:
\beq\label{dse}
\left\{\begin{array}{l}
\frac{1}{G(p^2)}-\frac{1}{G(\sigma_g^2)}=-\left(\Sigma(p^2)
-\Sigma(\sigma_g^2)\right)\\[.2cm]
\frac{1}{F(p^2)}-\frac{1}{F(\sigma_f^2)}=-\left(\Pi(p^2)
-\Pi(\sigma_f^2)\right),
\end{array}\right.
\eeq
where $\sigma_g$ and $\sigma_c$ are renormalization points for gluon and ghost propagator respectively. After truncation, the ghost self-energy and the gluon vacuum polarization are,
\beq
\begin{array}{rcl}
\Sigma(p^2)&=&N_c g_\mu^2\int T_0(p^2,q^2,r^2)G(q^2) F(r^2) \frac{d^d q}{(2\pi)^d},\\[.2cm]
\Pi(p^2)&=&\Pi^{2c}(p^2)+\Pi^{2g}(p^2)+\Pi^{4g}(p^2)
\end{array}
\eeq
where $g_\mu=\sqrt{4\pi \alpha(\mu)}$, $\mu$ is renormalization point, and separate contributions to vacuum polarization are
\beq
\begin{array}{rcl}\displaystyle
\Pi^{2c}(p^2)&=&\int  \frac{d^d q}{(2\pi)^d} M_0(p^2,q^2,r^2)G(q^2)G(r^2),\\[.2cm]
\Pi^{2g}(p^2)&=&\int  \frac{d^d q}{(2\pi)^4} Q_0(p^2,q^2,r^2)G(q^2)G(r^2),\\[.2cm]
\Pi^{4g}(p^2)&=&\int  \frac{d^d q}{(2\pi)^d} Q_4(p^2,q^2,r^2)\times \\ && \times G(q^2)^3F(q^2)^{3/4}
G(r^2)F(r^2)^{1/4},
\end{array}
\eeq
with kernels $M_0,Q_0,Q_4,T_0$ defined in\mycite{Bloch:2003yu}, eqs. (47, 48, 49, E.1). Here $x=p^2,y=q^2,z=r^2$, $r^2=(p-q)^2$.

We decompose the gluon and ghost form factors in a set of Chebyshev polynomials $T_i(z)$ on an interval $p^2\in (\epsilon,\sigma)$, and use a power-like Ansatz for the IR domain $p^2\in (
0,\epsilon)$:
\beq
F(z)=\left\{
\begin{array}{l}
\exp{\left(\sum_i f_i T_i(z)\right)}, z>\epsilon\\[0.3cm]
A z^{2\kappa},z<\epsilon
\end{array}\right.,
\eeq
and a similar decomposition is assumed for $G$,
where  $z=\log \frac{p^2}{\mu_0^2}$, $\mu_0$ is scale unit, continuity is imposed at $p^2=\epsilon$, renormalization points are chosen for gluon and ghost respectively, $\sigma_f=\sigma,\sigma_g=0$, normalization conditions $F(\sigma)=1$, $1/G(0)=0$; and from now on, $d=4$. For the two sets of coefficients we get algebraic equations, which are solved by Newton's method. Reliability of the method is supported by convergence of the sets of coefficients.

{\bf Choice of scale:}
Upon solving the equations, we construct running coupling out of the form factors from DSE
\beq\label{alphanp}
\alpha(p^2)=\alpha(\mu) F(p^2) G^2(p^2).
\eeq
This follows since we did not include the ghost-ghost-gluon vertex factor into the system of DSE, it is considered to be unity. Here $\mu$ is the renormalization scale for coupling, which is identical with the ultraviolet ~(UV) cutoff in our case: $\mu=\sigma$, and $F(\mu^2)=G(\mu^2)=1$

To relate the gauge coupling constant $\alpha$ with physics, one must introduce a scale. Physical scale is not fully defined in this problem, for we cannot actually compare pure gluodynamics to real-world QCD with fermions. Even so, we note that any object of dimension $k$ must be measured in units of $\mu_0^k$, where scale $\mu_0$ is one and the same for the whole problem. By comparing our (non-perturbative) $\alpha$ with the standard running coupling value\mycite{Amsler:2008zz} we obtain a rough estimate for our scale $\mu_0$
\beq
\alpha(\sqrt{s}=10 \mathrm{GeV})=0.18,\,\,\to \mu_0\sim 0.55 \mathrm{GeV}.
\eeq
For comparison, an estimate performed at $\sqrt{s}=m_\tau=1.77\mathrm{GeV},\,\, \alpha=0.35$ yields $\mu_0=0.8$.

{\bf From DSE to condensates:}
Our solutions of DSE confirm the results by Bloch in every feature; thus we proceed to applying them for obtaining the condensate. The form factors which we have obtained and the running coupling are true non-perturbative quantities  $F^{NP}(z),G^{NP}(z),\alpha(z)$. Now, in order to employ Eq.\myref{bilocal}, we need the difference between the perturbative and non-perturbative objects.

Bloch in\mycite{Bloch:2003yu} observes that he reproduced the leading-order PT results for coupling in the UV, but not the NLO. We confirm this by comparison of $\beta(\alpha)$  and anomalous dimension $\gamma(\alpha)$ for the propagator, see below, with their perturbative behaviour. We believe this is a consequence of the partial rather than full resummation of the perturbation series, since not all of the NLO diagrams emergent from the systematic expansion appear in DSE as solved here,  due to the truncation of the hierarchy of DSE.  In this sense we might call the present approximation resummed 1.5 loops. Resummed, since the partial resummation extends to all orders.     For possible improvements of our method, it would be advantageous to use DSE with vertex function included\mycite{Alkofer:2008bs}.

Given that the DSE solution is of infinite order in $\alpha$, but incomplete as of 2nd order, one must re-evaluate perturbation theory form factor, rendering it consistent with the {\it partial} resummation inherent in the truncated DSE. This must be always done since DSE is always in practical work a truncated system. Gluon form factor, normalized at $\mu$, is expected to be organized in the form:
\beq
F(p^2)=\exp\left(\int_{\alpha(\mu^2)}
^{\alpha(p^2)} \frac{b(\alpha)}{\alpha} d\alpha\right),
\eeq
where $b(\alpha(z))=\alpha(z)\gamma(\alpha(z))/\beta(\alpha(z))$, $z$, as before, is renormgroup time.
Functions $\beta(\alpha)$ and $\gamma(\alpha)$ are nonperturbative and given by:
\beq
\beta(\alpha(z))=\frac{\partial \alpha(z)}{\partial z},\qquad
\gamma(\alpha(z))=\frac{\partial \log F(z)}{\partial z}.
\eeq
In perturbation theory\mycite{Larin:1993tp} $b^{PT}(\alpha)=\frac{1}{\alpha}\sum_{i=0} b_i\alpha^i$, where $b_0^{PT}=\frac{13}{22},\,
b_1^{PT}=\frac{537}{968}, b_2^{PT}=\left(\frac{129791}{42592}-\frac{243 \zeta (3)}{176}\right)$. However instead of these $b_i^{PT}$ we must use results of a polynomial fit in the UV range of our ratio
$b(\alpha)=\frac{\alpha \gamma(\alpha)}{\beta(\alpha)}$:
\beq\label{fitb}
b(\alpha)\stackrel{\mathrm{Fit}}{\to} \sum_{i=0}^n \tilde{b}_i \alpha^i(z)=\tilde{b}^{PT}(\alpha).
\eeq
To obtain a reasonable precision,   $n=2$ suffices. Using $\beta,\gamma$ from DSE and having performed the fit Eq.\myref{fitb},  we apply the coefficients $\tilde{b}_i$ to build an ``effective'' perturbative expression for gluon form factor, which is to be afterwards subtracted from the NP expression:
\beq\label{ptprop}
F^{PT}(p^2)=\exp\left(\int_{\alpha(\mu^2)}
^{\alpha(p^2)} \frac{1}{\alpha} \sum_{i=0}\tilde{b}_i\alpha^i d\alpha\right).
\eeq

\begin{figure}[t!] \begin{center} \includegraphics[height = 4.5cm, width=8cm]{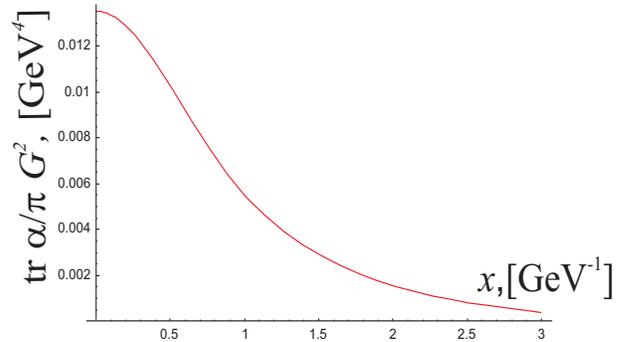} \caption{The nonlocal gluon condensate in $3+1$-dimensional $SU(3)$ from DSE approach, scale fixed at 10 GeV.} \label{FULLcondensate} \end{center} \end{figure}

We have now  got both, a non-perturbative propagator from solving DSE, and the perturbative propagator, Eq.\myref{ptprop}. Thus we can obtain the nonlocal condensate shown  in figure~\ref{FULLcondensate}. This is our main computational result. Our solution qualitatively satisfies the expectations which are due to the known UV and IR forms of the condensate, prescribed by OPE and lattice respectively. OPE predicts at small distances\mycite{Mikhailov:1992ug}
\beq
\mathcal{F}(x)=G_0^2(1-\lambda^2 x^2).
\eeq
Comparison with this short-distance behaviour is shown in  table~\ref{compar}. At large distances one can use~\mycite{gia}
\beq
\mathcal{F}(x)=Ce^{-ax},
\eeq
(to be used with caution, as some other predictions e.g. from instanton models\mycite{Dorokhov:1999ig}, exist).
It is instructive to inspect  numerical  convergence to the values of $G_0^2,\lambda^2,C,a$ shown in  table~\ref{converg}, as function of the number $n$  of polynomials involved. Results are stated for scale fixed at 10 GeV.
The remaining numerical error in the determination of the magnitude of  condensate  and correlation lengths can be eliminated by increasing the size of the basis, once there is merit in such an effort. Bloch\mycite{Bloch:2003yu} used $n=50$ in solving DSE. We estimate our current numerical error based on convergence shown in  table~\ref{converg} and show this in the bottom entries in  table~\ref{compar}.

{\bf Conclusion:}
We described here what we believe is a novel idea allowing to obtain  the non perturbative vacuum condensate, which has the capability of yielding both IR and  UV vacuum properties. This approach is less expensive than lattice simulations, and more universal than OPE.
\begin{table}
\centering
\caption{Comparison our results  with prior work.  ``SR'' denotes sum rules, $\tau$ means ``$\tau$ decay''. Results of our present Letter given in bold in the two last lines, denoted by $\dagger$ and $\ddagger$, for scale fixing points 1.8 GeV and 10 GeV respectively. Error bars are given in parentheses.}\label{compar}
\begin{tabular}{|l|l|l|l|l|l|l|}\hline
Method&Yr.&Ref.&$G_0^2$[$\mathrm{GeV}^4$]&
$\lambda^2$[$\mathrm{GeV}^2]$
&$C[\mathrm{GeV}^4]$&$a[\mathrm{GeV}$]\\  \hline
SR&'78&\cite{Shifman:1978bx}&.012&--&--&.--\\ \hline
OPE&'92&\cite{Mikhailov:1992ug}&--&.21&--&--\\  \hline
Lattice&'97&\cite{D'Elia:1997ne}&.015&--&.008&.6\\  \hline
Lattice&'02&\cite{D'Elia:2002ck}&--&--&.08&.8\\  \hline
SR&'02&\cite{Ioffe:2002ee}&.009(.007)&--&--&--\\ \hline
$\tau$&'02&\cite{Ioffe:2002ee}&.006(.012)&--&--&--\\ \hline\hline
\bf DSE&\bf '08&\bf $\dagger$&\bf .062(.022)&\bf 2.3(.6)&\bf .09(.03)&\bf  1.9(.4)\\ \hline
\bf DSE&\bf '08& \bf $\ddagger$& \bf .014(.005)&\bf1.1(.3)&\bf.020(.007)&\bf 1.3(.3)\\
 \hline \hline
\end{tabular}
\end{table}
\begin{table}
\centering
\caption{Convergence of condensates and correlation lengths with respect to the number of polynomials $n$.}\label{converg}
\begin{tabular}{|l|l|l|l|l|}\hline
$n$&$G_0^2$[$\mathrm{GeV}^4$]&
$\lambda^2$[$\mathrm{GeV}^2$]
&$C[\mathrm{GeV}^4]$&$a[\mathrm{GeV}]$\\ \hline
22 & 0.00834 & 1.04052 & 0.02671 & 2.26331\\ \hline
24 & 0.01068 & 0.64805 & 0.01854 &1.22284\\ \hline
26 & 0.00967 & 1.30711 & 0.01573 & 1.41736\\ \hline
28 & 0.01270 & 1.18162 & 0.01892 & 1.33049 \\ \hline
30 & 0.01352 & 1.12676 & 0.01977 & 1.28204 \\ \hline
\end{tabular}
\end{table}

Our main result is the demonstration that the solution of DSE allows determination of the full nonlocal gluon condensate $\langle: \frac{\alpha}{\pi}\frac{\tr}{N_c}G_{\mu\nu}(x)W(x,0)
G_{\mu\nu}(0)W(0,x):\rangle$ as a function of coordinate or momentum. From these one can get its fundamental properties: the value at zero separation and correlation range. Our results are obtained in Landau gauge, yet the procedure and thus the final answer are gauge-invariant. Our here presented method and results  suggest  that the effort required, especially including  Fermi (quark) fields, is greatly reduced compared to lattice method.

Moreover, in principle there seems to be no  obstacle to the study of
the far infrared limit, which relies of further development of DSE in this domain.
Our pure gauge theory results cannot be as yet used for further phenomenological developments due to scale uncertainty; however, the purpose of this paper was to demonstrate in principle how an {\it a priori} calculation of condensate is possible in the framework of DSE.

{\bf Acknowledgements:}
We are grateful to Jacques~Bloch for introducing us into the details of his DSE method. We thank A.~Bakulev, A.~Dorokhov and S.~Mikhailov for discussions. Numerical calculations were performed in part on the Computational Cluster of the Theoretical Division of INR RAS, Moscow.  A.Z.  would like to thank D.V.~Shirkov and A.S.~Gorsky for their kind advice.
We thank P.D. Dr. Peter Thirolf and  Prof. Dr. D. Habs, Director of  the Cluster of Excellence in Laser Physics -- Munich-Center for Advanced Photonics  (MAP) for their hospitality
in Garching while much of this research was carried out.
This research was supported  in part by RFBR Grant 07-01-00526, by the DFG--LMUexcellent  grant, and
by a grant from: the U.S. Department of Energy  DE-FG02-04ER4131.

\bibliography{dse-letM}

\begin{thebibliography}{19}
\expandafter\ifx\csname natexlab\endcsname\relax\def\natexlab#1{#1}\fi
\expandafter\ifx\csname bibnamefont\endcsname\relax
  \def\bibnamefont#1{#1}\fi
\expandafter\ifx\csname bibfnamefont\endcsname\relax
  \def\bibfnamefont#1{#1}\fi
\expandafter\ifx\csname citenamefont\endcsname\relax
  \def\citenamefont#1{#1}\fi
\expandafter\ifx\csname url\endcsname\relax
  \def\url#1{\texttt{#1}}\fi
\expandafter\ifx\csname urlprefix\endcsname\relax\def\urlprefix{URL }\fi
\providecommand{\bibinfo}[2]{#2}
\providecommand{\eprint}[2][]{\url{#2}}

\bibitem[{\citenamefont{Shifman et~al.}(1979)\citenamefont{Shifman, Vainshtein,
  and Zakharov}}]{Shifman:1978bx}
\bibinfo{author}{\bibfnamefont{M.~A.} \bibnamefont{Shifman}},
  \bibinfo{author}{\bibfnamefont{A.~I.} \bibnamefont{Vainshtein}},
  \bibnamefont{and} \bibinfo{author}{\bibfnamefont{V.~I.}
  \bibnamefont{Zakharov}}, \bibinfo{journal}{Nucl. Phys.}
  \textbf{\bibinfo{volume}{B147}}, \bibinfo{pages}{385} (\bibinfo{year}{1979}).

\bibitem[{\citenamefont{Mikhailov}(1993)}]{Mikhailov:1992ug}
\bibinfo{author}{\bibfnamefont{S.~V.} \bibnamefont{Mikhailov}},
  \bibinfo{journal}{Phys. Atom. Nucl.} \textbf{\bibinfo{volume}{56}},
  \bibinfo{pages}{650} (\bibinfo{year}{1993}).

\bibitem[{\citenamefont{Lavelle and Schaden}(1988)}]{Lavelle:1988eg}
\bibinfo{author}{\bibfnamefont{M.~J.} \bibnamefont{Lavelle}} \bibnamefont{and}
  \bibinfo{author}{\bibfnamefont{M.}~\bibnamefont{Schaden}},
  \bibinfo{journal}{Phys. Lett.} \textbf{\bibinfo{volume}{B208}},
  \bibinfo{pages}{297} (\bibinfo{year}{1988}).

\bibitem[{\citenamefont{Rafelski}(1998)}]{Rafelski:1998tc}
\bibinfo{author}{\bibfnamefont{J.}
~\bibnamefont{Rafelski}}
%
in {\it Frontier Tests of QED and Physics of the Vacuum},
E. Zavattini {\it et al} eds., (Heron Press, Sofia 1998) pp 425-439; and 
%
 {\it Quantum Chromodynamics} Paris, France 1--6 June 1998,
World Scientific, H. Fried and B. Muller, eds., pp 208-223, 
\eprint{hep-ph/9806389}.

\bibitem[{\citenamefont{Bloch}(2003)}]{Bloch:2003yu}
\bibinfo{author}{\bibfnamefont{J.~C.~R.} \bibnamefont{Bloch}},
  \bibinfo{journal}{Few Body Syst.} \textbf{\bibinfo{volume}{33}},
  \bibinfo{pages}{111} (\bibinfo{year}{2003}), \eprint{hep-ph/0303125}.

\bibitem[{\citenamefont{Dosch and Simonov}(1988)}]{Dosch:1988ha}
\bibinfo{author}{\bibfnamefont{H.~G.} \bibnamefont{Dosch}} \bibnamefont{and}
  \bibinfo{author}{\bibfnamefont{Y.~A.} \bibnamefont{Simonov}},
  \bibinfo{journal}{Phys. Lett.} \textbf{\bibinfo{volume}{B205}},
  \bibinfo{pages}{339} (\bibinfo{year}{1988}).

\bibitem[{\citenamefont{Namyslowski}()}]{Namyslowski:1991wu}
\bibinfo{author}{\bibfnamefont{J.~M.} \bibnamefont{Namyslowski}},
  \bibinfo{note}{Preprint IFT-11-91}, Warsaw Univ.

\bibitem[{\citenamefont{Hauck et~al.}(1998)\citenamefont{Hauck, von Smekal, and
  Alkofer}}]{Hauck:1996sm}
\bibinfo{author}{\bibfnamefont{A.}~\bibnamefont{Hauck}},
  \bibinfo{author}{\bibfnamefont{L.}~\bibnamefont{von Smekal}},
  \bibnamefont{and} \bibinfo{author}{\bibfnamefont{R.}~\bibnamefont{Alkofer}},
  \bibinfo{journal}{Comput. Phys. Commun.} \textbf{\bibinfo{volume}{112}},
  \bibinfo{pages}{149} (\bibinfo{year}{1998}), \eprint{hep-ph/9604430}.

\bibitem[{\citenamefont{Fischer and Alkofer}(2003)}]{Fischer:2003rp}
\bibinfo{author}{\bibfnamefont{C.~S.} \bibnamefont{Fischer}} \bibnamefont{and}
  \bibinfo{author}{\bibfnamefont{R.}~\bibnamefont{Alkofer}},
  \bibinfo{journal}{Phys. Rev.} \textbf{\bibinfo{volume}{D67}},
  \bibinfo{pages}{094020} (\bibinfo{year}{2003}), \eprint{hep-ph/0301094}.

\bibitem[{\citenamefont{Fischer}(2008)}]{Fischer:2008ic}
\bibinfo{author}{\bibfnamefont{C.~S.} \bibnamefont{Fischer}}
  (\bibinfo{year}{2008}), \eprint{arXiv:0810.2526}.

\bibitem[{\citenamefont{Fischer et~al.}(2008)\citenamefont{Fischer, Maas, and
  Pawlowski}}]{Fischer:2008yv}
\bibinfo{author}{\bibfnamefont{C.~S.} \bibnamefont{Fischer}},
  \bibinfo{author}{\bibfnamefont{A.}~\bibnamefont{Maas}}, \bibnamefont{and}
  \bibinfo{author}{\bibfnamefont{J.~M.} \bibnamefont{Pawlowski}}
  (\bibinfo{year}{2008}), \eprint{arXiv:0812.2745}.

\bibitem[{\citenamefont{Alkofer et~al.}(2008)\citenamefont{Alkofer, Fischer,
  Huber, Llanes-Estrada, and Schwenzer}}]{Alkofer:2008bs}
\bibinfo{author}{\bibfnamefont{R.}~\bibnamefont{Alkofer}},
  \bibinfo{author}{\bibfnamefont{C.~S.} \bibnamefont{Fischer}},
  \bibinfo{author}{\bibfnamefont{M.~Q.} \bibnamefont{Huber}},
  \bibinfo{author}{\bibfnamefont{F.~J.} \bibnamefont{Llanes-Estrada}},
  \bibnamefont{and} \bibinfo{author}{\bibfnamefont{K.}~\bibnamefont{Schwenzer}}
  (\bibinfo{year}{2008}), \eprint{arXiv:0812.2896}.

\bibitem[{\citenamefont{Amsler et~al.}(2008)}]{Amsler:2008zz}
\bibinfo{author}{\bibfnamefont{C.}~\bibnamefont{Amsler}} \bibnamefont{et~al.}
  (\bibinfo{collaboration}{Particle Data Group}), \bibinfo{journal}{Phys.
  Lett.} \textbf{\bibinfo{volume}{B667}}, \bibinfo{pages}{1}
  (\bibinfo{year}{2008}).

\bibitem[{\citenamefont{Larin and Vermaseren}(1993)}]{Larin:1993tp}
\bibinfo{author}{\bibfnamefont{S.~A.} \bibnamefont{Larin}} \bibnamefont{and}
  \bibinfo{author}{\bibfnamefont{J.~A.~M.} \bibnamefont{Vermaseren}},
  \bibinfo{journal}{Phys. Lett.} \textbf{\bibinfo{volume}{B303}},
  \bibinfo{pages}{334} (\bibinfo{year}{1993}), \eprint{hep-ph/9302208}.

\bibitem[{\citenamefont{Di~Giacomo et~al.}(2002)\citenamefont{Di~Giacomo,
  Dosch, Shevchenko, and Simonov}}]{gia}
\bibinfo{author}{\bibfnamefont{A.}~\bibnamefont{Di~Giacomo}},
  \bibinfo{author}{\bibfnamefont{H.~G.} \bibnamefont{Dosch}},
  \bibinfo{author}{\bibfnamefont{V.~I.} \bibnamefont{Shevchenko}},
  \bibnamefont{and} \bibinfo{author}{\bibfnamefont{Y.~A.}
  \bibnamefont{Simonov}}, \bibinfo{journal}{Phys. Rept.}
  \textbf{\bibinfo{volume}{372}}, \bibinfo{pages}{319} (\bibinfo{year}{2002}),
  \eprint{hep-ph/0007223}.

\bibitem[{\citenamefont{Dorokhov et~al.}(2000)\citenamefont{Dorokhov,
  Esaibegian, Maximov, and Mikhailov}}]{Dorokhov:1999ig}
\bibinfo{author}{\bibfnamefont{A.~E.} \bibnamefont{Dorokhov}},
  \bibinfo{author}{\bibfnamefont{S.~V.} \bibnamefont{Esaibegian}},
  \bibinfo{author}{\bibfnamefont{A.~E.} \bibnamefont{Maximov}},
  \bibnamefont{and} \bibinfo{author}{\bibfnamefont{S.~V.}
  \bibnamefont{Mikhailov}}, \bibinfo{journal}{Eur. Phys. J.}
  \textbf{\bibinfo{volume}{C13}}, \bibinfo{pages}{331} (\bibinfo{year}{2000}),
  \eprint{hep-ph/9903450}.

\bibitem[{\citenamefont{D'Elia et~al.}(1997)\citenamefont{D'Elia, Di~Giacomo,
  and Meggiolaro}}]{D'Elia:1997ne}
\bibinfo{author}{\bibfnamefont{M.}~\bibnamefont{D'Elia}},
  \bibinfo{author}{\bibfnamefont{A.}~\bibnamefont{Di~Giacomo}},
  \bibnamefont{and}
  \bibinfo{author}{\bibfnamefont{E.}~\bibnamefont{Meggiolaro}},
  \bibinfo{journal}{Phys. Lett.} \textbf{\bibinfo{volume}{B408}},
  \bibinfo{pages}{315} (\bibinfo{year}{1997}), \eprint{hep-lat/9705032}.

\bibitem[{\citenamefont{D'Elia et~al.}(2003)\citenamefont{D'Elia, Di~Giacomo,
  and Meggiolaro}}]{D'Elia:2002ck}
\bibinfo{author}{\bibfnamefont{M.}~\bibnamefont{D'Elia}},
  \bibinfo{author}{\bibfnamefont{A.}~\bibnamefont{Di~Giacomo}},
  \bibnamefont{and}
  \bibinfo{author}{\bibfnamefont{E.}~\bibnamefont{Meggiolaro}},
  \bibinfo{journal}{Phys. Rev.} \textbf{\bibinfo{volume}{D67}},
  \bibinfo{pages}{114504} (\bibinfo{year}{2003}), \eprint{hep-lat/0205018}.

\bibitem[{\citenamefont{Ioffe}(2003)}]{Ioffe:2002ee}
\bibinfo{author}{\bibfnamefont{B.~L.} \bibnamefont{Ioffe}},
  \bibinfo{journal}{Phys. Atom. Nucl.} \textbf{\bibinfo{volume}{66}},
  \bibinfo{pages}{30} (\bibinfo{year}{2003}), \eprint{hep-ph/0207191}.

\end{thebibliography}
\end{document}